\documentstyle[aps,pre,epsfig,12pt]{revtex}
\draft

\begin{document}
\title{Density Functional Theory Studies of Magnetically Confined 
Fermi Gas\footnote{Supported by the National Natural Science Foundation of 
China for the Distinguished 
Young Scholars, and Shanghai Science and Technology Committee.}
}
\author{ CHEN Yu-Jun,  MA Hong Ru \\
Department of Applied Physics, Shanghai Jiao Tong University, Shanghai 200240}
\maketitle

\begin{abstract}
   A theory is developed for magnetically confined Fermi gas at low
temperature based on the density functional theory. 
The theory is illustrated by numerical calculation of
density distributions of Fermi atoms $^{40}$K with parameters
according to   DeMarco 
and Jin's experiment[Science, 285(1999)1703].  
Our results are in good  agreement with 
the experiment. To check the theory, we also performed calculations
using our theory at high temperature and compared very well to
the result of classical limit.
\end{abstract}

In a recent publication, DeMarco and Jin reported their work 
on Fermi atoms at low temperature in confined geometries. They 
employed an evaporative cooling 
strategy to cool a magnetically trapped gas of $7\times 10^5$ ${}^{40}$K 
atoms to $0.5$ of the Fermi temperature $T_F$. 
An Ioffe-Pritchard-type magnetic trap\cite{Gott} provides a 
cylindrically symmetric, harmonic potential with an axial frequency of 
$\omega _z=2\pi \times 19.5$ Hz and a variable radial frequency. The 
radial frequency, as well as the minimum magnetic field, can be smoothly 
varied from $\omega _r=2\pi \times 44$ Hz to $2\pi \times 370$ Hz by 
changing the current in a pair of Helmholtz bias coils. 
In the experiment presented by DeMarco and Jin,
a single-component gas of the  $F=9/2$,
$m_F=9/2$ atoms is produced, where $F$ is the total atomic spin and $m_F$
is its magnetic quantum number. The time-of-flight image was taken by suddenly 
switching off the current that provides the magnetic trapping field, 
which allowed the gas to expand freely for 
$15$ to $20$ ms. The absorption shadow, generated by illumination of the 
expanded gas, was imaged onto a charged-coupled device array.
They detected the emergency of quantum degeneracy in a trapped gas 
of Fermionic atoms and observed a non-classical momentum distribution 
and found that the total energy of the gas is larger than the classical 
expectation. They observed not only the momentum distribution but also the 
confined gas itself, which approaches a fixed size as 
$T$ approaches zero.

In this letter, we present a general theory of the low-temperature 
density profile of an ideal Fermi gas trapped within an arbitrary 
potential well. Then we give the results of numerical calculation to 
compare to the experiment. The approach taken here is based on the 
density functional theory (DFT)\cite{Mermin,Gross}, which was first
introduced by Kohn and Hohenberg\cite{Kohn}
 in the context of ground state energy 
of quantum systems. It was developed and extended to excited states and
to finite temperatures by some researchers and has become the effective 
first principle calculational method for 
the electronic and structural properties of a large variety of 
condensed matter systems. Considering a single-component system 
within an external potential at 
$T\ne 0K$, the DFT asserts that the true density distribution of 
a system in an external field $V({\mathbf r})$ is the one that 
leads to the minimum of the following functional,
\begin{equation} \label{eq:001}
F[n({\mathbf r})]=F_0[n({\mathbf r})]+
\int {[V({\mathbf r})-\mu ]\cdot 
n({\mathbf r})d{\mathbf r}},
\end{equation}
and the minimum is the 
the Helmholtz free energy of the system.
Here $\mu $ is the chemical potential in the above expression. While 
$F_0[n({\mathbf r})]$ is a temperature dependent functional of 
$n({\mathbf r})$ only, the dependence of $F_0[n({\mathbf r})]$ on external
potential $V({\mathbf r})$ is only through the dependence of 
$n({\mathbf r})$ on $V({\mathbf r})$. This is a very strong statement
which means that if we know the functional form of $F_0[n({\mathbf r})]$
we can get all the physical quantities of interest.
 For a given external potential, the 
functional derivative of 
$F[n({\mathbf r})]$ with respect to 
$n({\mathbf r})$ should equals to zero:
\begin{equation} \label{eq:002}
{{\delta F[n({\mathbf r})]} \over 
{\delta n({\mathbf r})}}=0.
\end{equation}          
This gives an equation for $n({\mathbf r})$, and when
 $n({\mathbf r})$ is obtained from the solution of Eq.~(\ref{eq:002}),
we substitute it back to (\ref{eq:001}) to get the free energy and other 
quantities can be obtained simply by differential. 
It is hard to get the functional form of $F_0[n({\mathbf r})]$,
various approximations are employed in practical calculations. One
of the commonly used approximation is the local density approximation(LDA)
which usually gives excellent results to real systems. We use here
the LDA in 
our theory of confined Fermi systems. In this approximation the functional
$F_0[n({\mathbf r})]$ is assumed to be 
\begin{equation}\label{eq:003}
F_0[n({\mathbf r})]=\int d{\mathbf r} f_0(n({\mathbf r})).
\end{equation}
And we assume that the free energy density $f_0(n({\mathbf r}))$
has also the same functional form as the free energy of an 
 uniform system of density $n$, i.e, 
$f_0(n)$ is the Helmholtz free energy density of an 
uniform system with density $n$.
Then we can obtain
\begin{equation} \label{eq:004}
{{\delta F[n({\mathbf r})]} \over {\delta n({\mathbf r})}}
={{\partial f_0(n({\mathbf r}))} \over {\partial n({\mathbf r})}}+
[V({\mathbf r})-\mu ].
\end{equation}
According to Eqs.~(\ref{eq:002}) and (\ref{eq:004}), we have 
\begin{equation} \label{eq:005}
{{\partial f_0(n)} \mathord{\left/ {\vphantom {{\partial f_0(n)}
 {\partial n}}} \right. \kern-\nulldelimiterspace} {\partial n}}
=\mu -V({\mathbf r}).
\end{equation}
For the ideal Fermi gas, we know\cite{Huang}
\begin{equation} \label{eq:006}
n={1 \over {\lambda ^3}}f_{{3 \mathord{\left/ {\vphantom {3 2}} \right.
 \kern-\nulldelimiterspace} 
2}}(z),
\end{equation}
and                           
\begin{equation} \label{eq:007}
f_0=-{{kT} \over {\lambda ^3}}f_{{5 \mathord{\left/ {\vphantom {5 2}} 
\right. \kern-\nulldelimiterspace} 2}}(z)+n\,kT\log z,
\end{equation}
here
\begin{eqnarray*}
f_{{5 \mathord{\left/ {\vphantom {5 2}} \right. \kern-\nulldelimiterspace} 
2}}(z)&=&{4 \over {\sqrt \pi }}\int_0^\infty  
{dx\cdot x^2\log (1+ze^{-x^2})}, \\
f_{{3 \mathord{\left/ {\vphantom {3 2}} \right. \kern-\nulldelimiterspace}
 2}}(z)&=&z{\partial  \over 
{\partial z}}f_{{5 \mathord{\left/ {\vphantom {5 2}} \right. 
\kern-\nulldelimiterspace} 2}}(z),
\end{eqnarray*}
where 
$k$ is Boltzmann's constant, 
$\lambda =\sqrt {{{2\pi \hbar ^2} \mathord{\left/ {\vphantom {{2\pi \hbar ^2} 
{mkT}}} \right. \kern-\nulldelimiterspace} {mkT}}}$ the thermal wave 
length,
$z=e^{\beta \mu _0}$ the fugacity, and  
$\mu _0$ the chemical potential of the uniform system. So we get
\begin{equation} \label{eq:008}
{{\partial f_0(n)} \mathord{\left/ {\vphantom {{\partial f_0(n)}
 {\partial n}}} \right. \kern-\nulldelimiterspace}
 {\partial n}}=kT\log z.
\end{equation}
The value of 
$\mu$ is determined from the normalization of 
$n({\mathbf r},\mu )$:
\begin{equation} \label{eq:009}
N=\int {n({\mathbf r},\mu )d{\mathbf r}}
\end{equation}
where we have indicated the 
$\mu $ dependence of 
$n$ which follows from Eq.~(\ref{eq:005}). Once having determined 
$\mu $ from Eq.~(\ref{eq:009}) we substitute it back into Eq.~(\ref{eq:006}) to get 
$n({\mathbf r})$. Oliva has developed an approximate form of the Helmholtz free energy 
of the uniform system for different regimes of density\cite{Oliva}.
However, in our theory such an expression of 
$f_0$ with respect to $n$ is not required.

Now we use the confining potential well, according to the experiment 
by DeMarco and Jin, to calculate the density distributions
of Fermi atoms.

We consider a harmonic potential form
$$V(r,z)={1 \over 2}m\omega _r^2r^2+{1 \over 2}m\omega _z^2z^2$$
which is cylindrically symmetric, where the axial frequency 
$\omega _z=2\pi \times 19.5$Hz and the radial frequency 
$\omega _r=2\pi \times 137$Hz. 
The potential is  independent of the azimuth angle. We have used the parameters
given by the experiment setup, which is $N=7\times 10^5$ and 
$T={{T_F} \mathord{\left/ {\vphantom {{T_F} 2}} \right. 
\kern-\nulldelimiterspace} 2}$, with
$T_F={{\hbar (6\omega _z\omega _r^2N)^{{1 \mathord{\left/ {\vphantom {1 3}}
 \right. \kern-\nulldelimiterspace} 3}}} 
\mathord{\left/ {\vphantom {{\hbar (6\omega _z\omega _r^2N)^{{1 
\mathord{\left/ {\vphantom {1 3}} \right. \kern-\nulldelimiterspace} 3}}} k}} 
\right. \kern-\nulldelimiterspace} k}$\cite{Butts,Silvera}. The value of 
$T_F$ is 
$0.6\mu K$ for a million atoms in the 
$\omega _r=2\pi \times 137$Hz trap.
 
\begin{figure}[h]
\begin{minipage}{0.45\linewidth}
\epsfig{file=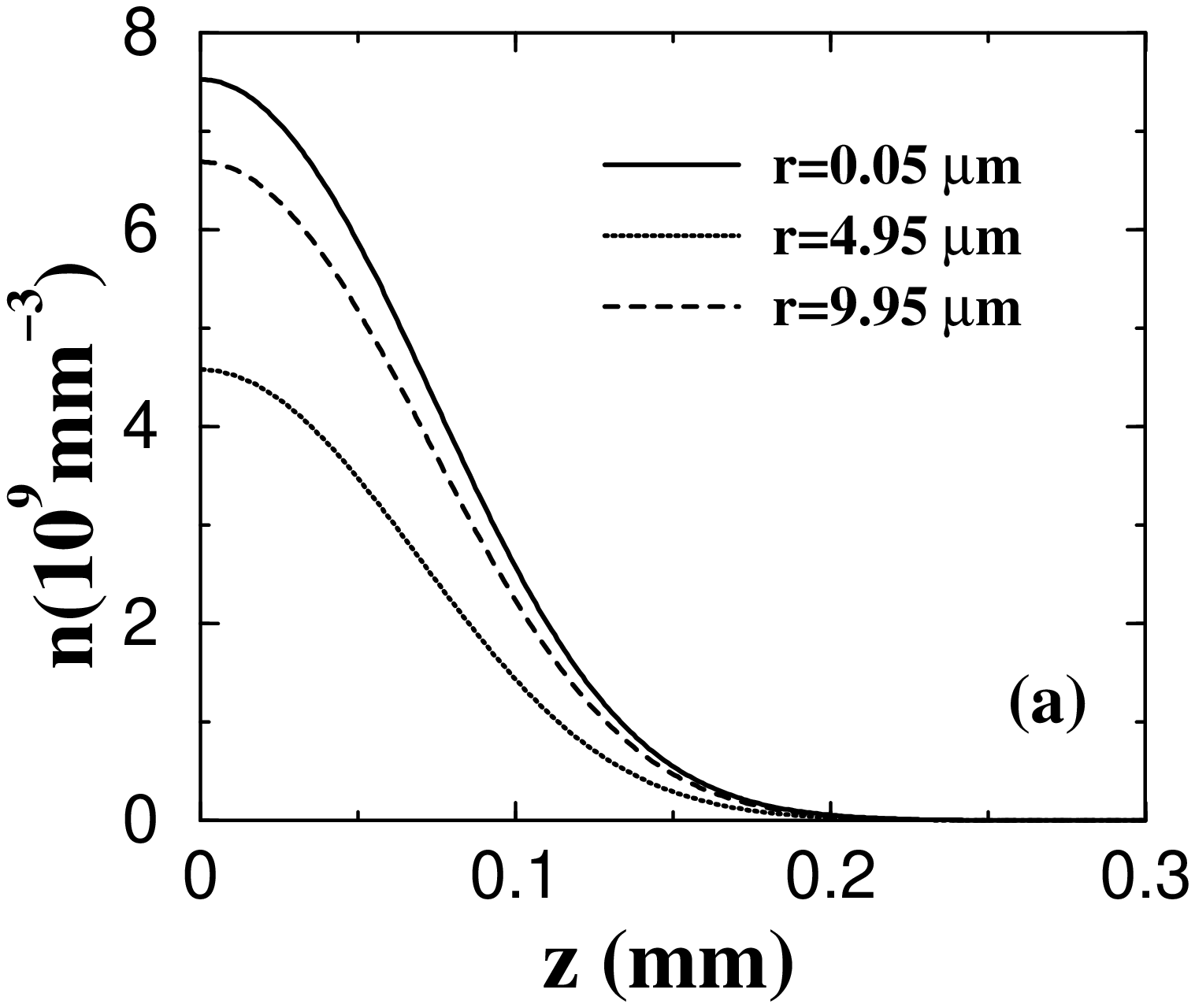,width=0.9\linewidth}
\end{minipage}
\begin{minipage}{0.45\linewidth}
\epsfig{file=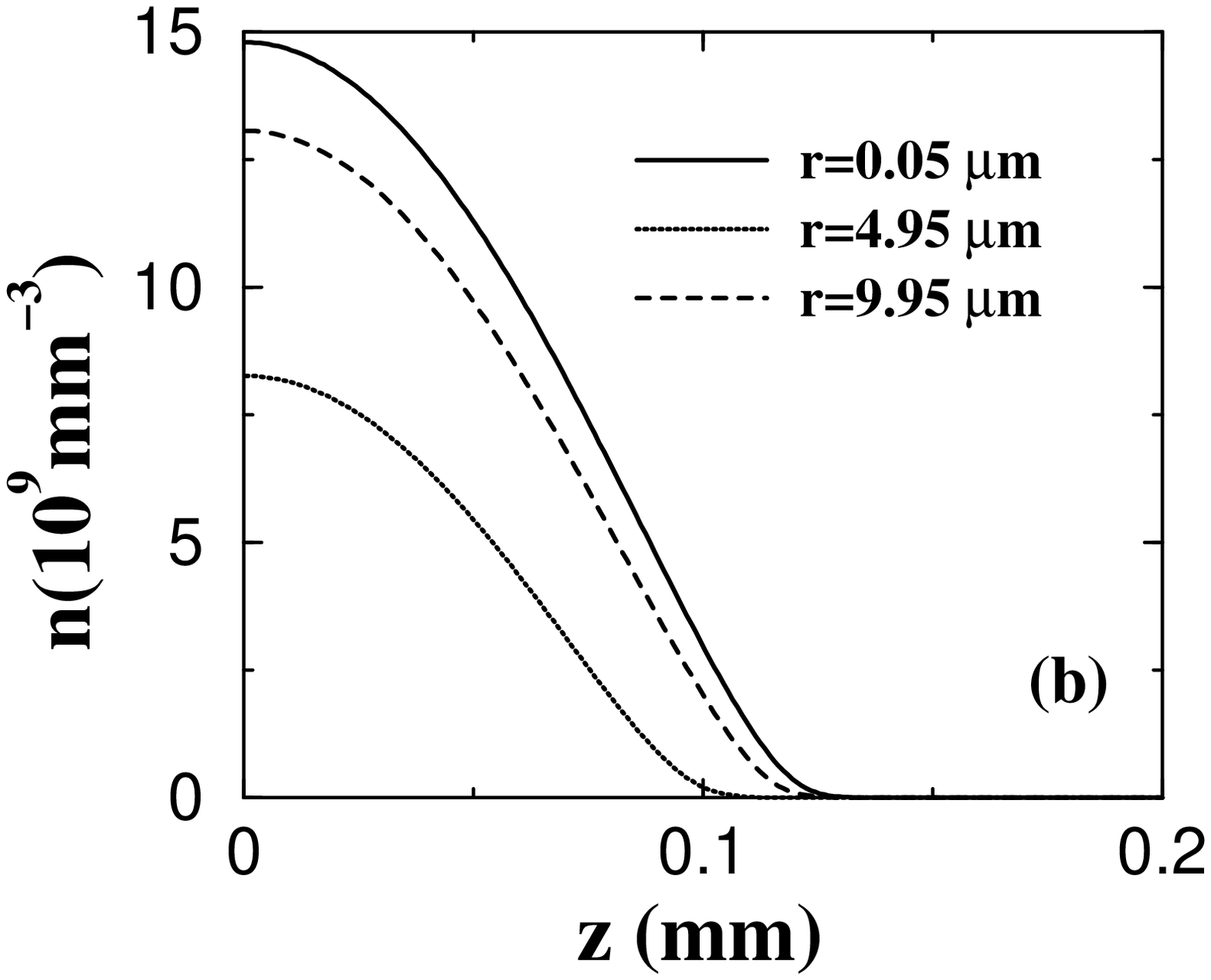,width=0.9\linewidth}
\end{minipage}
\caption{Atomic density as function of $z$ with different values of $r$
for $^{40}$K atoms under magnetic confinement.
Total number of atoms $N=7\times 10^5$, (a) $T=0.5 T_F$, (b) $T=0.05T_F$,
 with $T_F=0.6\mu K$.}
\end{figure}
\begin{figure}[h]
\begin{minipage}{0.45\linewidth}
\epsfig{file=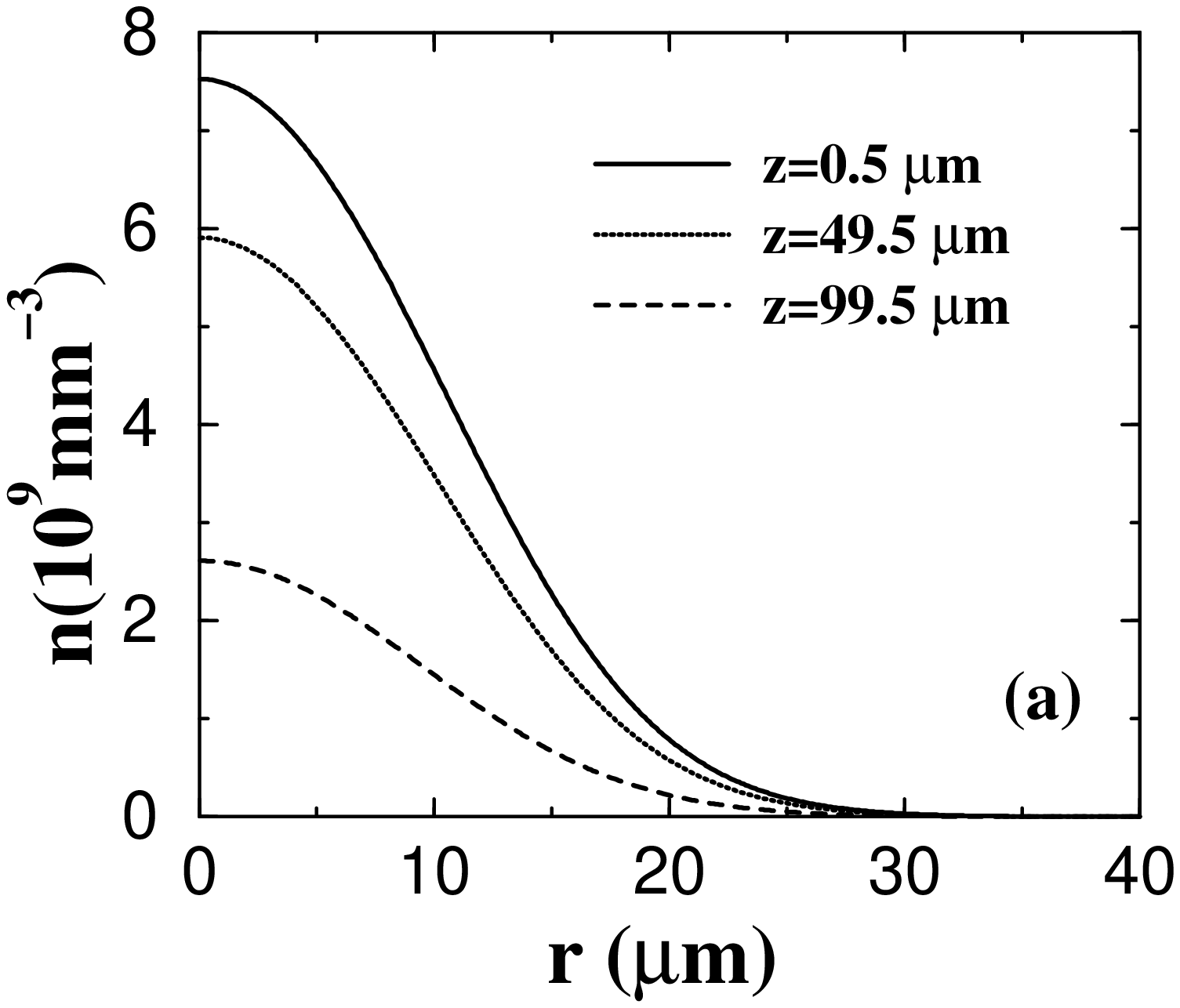,width=0.9\linewidth}
\end{minipage}
\begin{minipage}{0.45\linewidth}
\epsfig{file=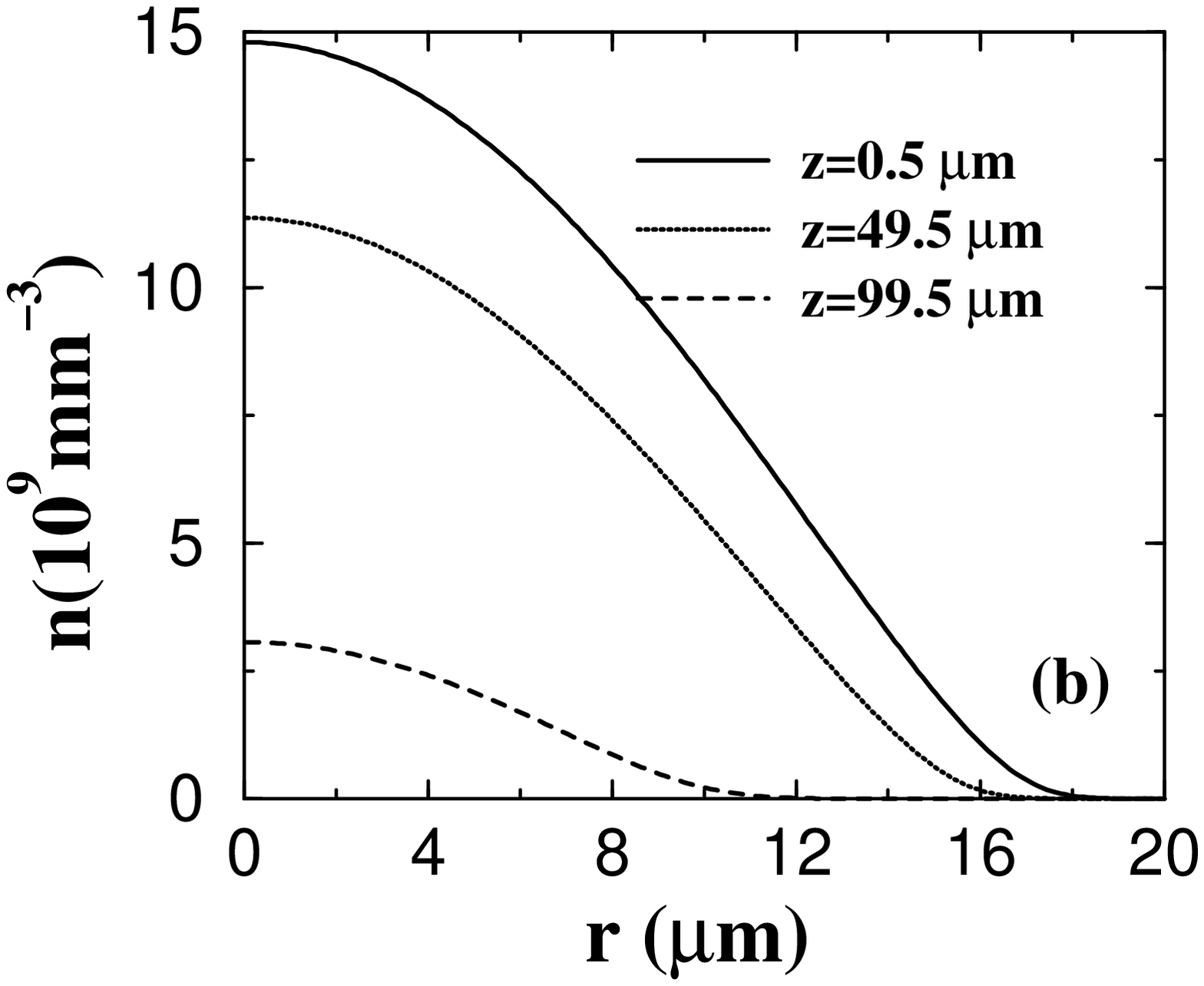,width=0.9\linewidth}
\end{minipage}
\caption{Atomic density as function of $r$ with different values of $z$
for $^{40}$K atoms under magnetic confinement. Other parameters are the same
as Fig. 1.}
\end{figure}

The particle density distributions are calculated by using the present method. 
Figures 1 and 2 are plots of the density as function
of $z$ and  $r$ for two different temperatures.
As observed in the experiment that the confined gas 
approaches a fixed size, we 
calculated the root-mean-square radius of the system and get
$r_{RMS}=0.7R_F$ at $T=0.5 T_F$ and $r_{RMS}=0.5R_F$ at low temperature
$T=0.05T_F$, where 
$R_F=\sqrt {{{2kT_F} \mathord{\left/ {\vphantom {{2kT_F} {m\omega _r^2}}} 
\right. \kern-\nulldelimiterspace} {m\omega _r^2}}}$\cite{Butts,Silvera}. 
These results are coincident with the experiment, where 
they obtained 
$r_{RMS}=0.6R_F$ as $T$ approaches zero and 
$r_{RMS}=0.9R_F$ at 
${T \mathord{\left/ {\vphantom {T {T_F}}} \right. \kern-\nulldelimiterspace} 
{T_F}}=0.5$\cite{DeMarco} . The difference between theory and experiment 
may due to 
the systematic uncertainty in 
${T \mathord{\left/ {\vphantom {T {T_F}}} \right. \kern-\nulldelimiterspace}
 {T_F}}$ within the experiment. To check the validity of our method we 
calculated the case for 
${T \mathord{\left/ {\vphantom {T {T_F}}} \right. \kern-\nulldelimiterspace}
 {T_F}}=2$ with other conditions unchanged. At high temperature (i.e., in the 
classical limit 
$T>>T_F$), we have\cite{Butts}:
\begin{equation} \label{eq:010}
\mu =-kT\log [6({T \over {T_F}})^3]
\end{equation}      
In  our calculation the chemical potential of the system is
$\mu =-5.86\times 10^{-29}$ J at 
${T \mathord{\left/ {\vphantom {T {T_F}}} \right. \kern-\nulldelimiterspace} 
{T_F}}=2$, while  
Eq. (10) gives  $-5.92 \times 10^{-29}$ J, which is satisfactory. 
The comparison of density distributions of our theory with the classical 
statistics can be seen in Fig. 3. 

\begin{figure}[h]
\epsfig{file=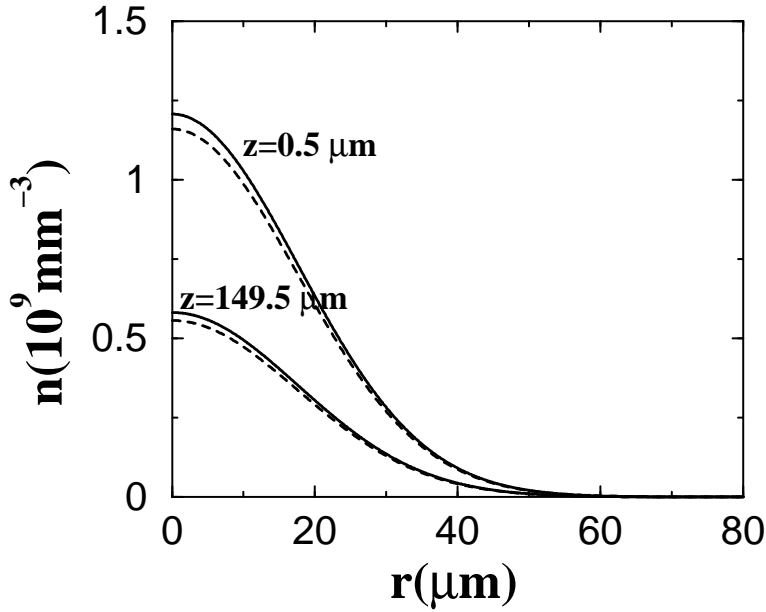,width=0.6\linewidth}
\caption{Atomic density as function of $r$ with different values of $z$
for $^{40}$K atoms under magnetic confinement. 
Parameters are the same as Fig. 1 
except $T=2T_F$. The solid lines are calculations from DFT as described 
in this work and dashed lines are from the Boltzmann distributions.}
\end{figure}

    In summary we calculated the physical
properties of the Fermi gas confined within a 
potential well at none-zero temperature. This method is based on the 
density functional theory under local density approximation.
The case of $^{40}$K atoms in a magnetic confinement is calculated 
and the result agrees with recent experiment. Here the interaction between
atoms is completely neglected and the study of effect of interaction on 
the confined Fermi atomic gas  is underway. Further studies of quantum 
degenerate Fermi gases in different confinement and interactions
are also the important and interesting subjects.


\begin{thebibliography}{100}
\bibitem{DeMarco} B. DeMarco and D. S. Jin, Science, {\bf 285}(1999)1703.
\bibitem{Gott} Y. V. Gott, M. S. Ioffe and V. G. Tel'kovski, Nucl. Fusion, 
suppl.(1962)1045; 1284; D. E. Pritchard, Phys. 
Rev. Lett. {\bf 51}(1983)1336.
\bibitem{Huang} See, e.g., Kerson Huang, Statistical Mechanics 
(John Wiley \& Sons, Inc., New York, 1987).
\bibitem{Mermin}N. D. Mermin, Phys. Rev. A {\bf 137}(1965)1141.
\bibitem{Gross}E. K. U. Gross and R. M. Dreizler, Density Functional Theory 
(Plenum Press, New York, 1995).
\bibitem{Kohn} P. Hohenberg and W. Kohn, Phys. Rev. B {\bf 136}(1964)864.
\bibitem{Oliva}J. Oliva, Phys. Rev. B {\bf 39}(1989)4204.
\bibitem{Butts}D. A. Butts and  D. S. Rokhsar, Phys. Rev. A {\bf 55}(1997)4346.
\bibitem{Silvera}I. F. Silvera and  J. T. M. Walraven, J. Appl. Phys. 
{\bf 52}(1981)2304.
\end{thebibliography}
\end{document}